\documentclass[11pt]{amsart}
\numberwithin{equation}{section}
\usepackage{graphicx}
\usepackage{hyperref}
\begin{document}

\title{One thing that general relativity says about photons in matter}
\author{S. Antoci and L. Mihich}
\address{Dipartimento di Fisica ``A. Volta'', Via Bassi 6, Pavia, Italy}
\email{Antoci@fisav.unipv.it}


\keywords{General relativity, electrodynamics, geometrical optics, light quanta}%

\begin{abstract}
Let us abandon for a moment the strict epistemological standpoint
of quantum field theory, that eventually comes to declare
nonsensical any question about the photon posed outside the
quantum theoretical framework. We can then avail of the works by
Whittaker et al. and by Synge about the particle and the wave
model of the photon in the vacuum of general relativity. We can
also rely on important results found by Gordon and by Pham Mau
Quan: thanks to Gordon's discovery of an effective metric these
authors have been able to reduce to the vacuum case several
problems of the electromagnetic theory of dielectrics.\par The
joint use of these old findings allows one to conclude that a
quantum theoretical photon in an isotropic dielectric has a
classical simile only if the dielectric is also homogeneous.
\end{abstract}
\maketitle

\section{Introduction}
When writing a paper with the title given above, one is fully
aware of the fact that it will be judged very differently by
readers with different epistemological inclinations. Imagine that
we strictly embrace what Schr\"odinger, in a memorable paper
\cite{ref:Schroedinger1935}, dubbed ``Der bewu{\ss}te Wechsel des
erkenntnistheo\-retischen Standpunkts'', and thereby ``wir haben
unsere naiv-realistische Unschuld verloren''. Photons are quantum
theoretical entities, and the positivistic credo currently
attached to quantum theory allows us to ask only questions that
deal with quantities that can be observed, at least in principle,
according to the quantum framework that we have adopted. Hence
there is no room for the good old ``Physik der Modelle'' of the
present writing, despite the fundamental role that such a kind of
physics has played in fostering the very birth of quantum
mechanics and of quantum field theory. For us, this sort of
physics shall be simply incompetent to say anything on the
subject, for it lies outside the allowed conceptual framework.\par
If, on the contrary, we still have not yet completely lost our
ingenuous-realistic innocence, we may feel relieved if the problem
that we should tackle quantum mechanically has a classical simile.
In that case, we can hope to rely on some clear-cut argument of
classical origin for guiding our steps if the quantum mechanical
apparatus, despite its programmatic self-reliance, for some reason
turns out to be inept at producing an answer whatsoever.\par This
lamentable occurrence happens when one attempts to deal with
quantum optics in dielectric media. As recently stressed
\cite{ref:Miglietta1998}, the usual presentations rely on {\it ad
hoc} effective Hamiltonians drawn in some way or other from the
classical electromagnetism of continua, thereby avoiding the
admittedly very hard, but necessary task of {\it deriving} an
effective Hamiltonian from a fully quantum description of the
whole system consisting of matter plus field. No wonder then, if
after more than five decades from the first attempt at producing
``phenomenological'' photons in homogeneous, isotropic matter
\cite{ref:Jauch1948}, \cite{ref:Nagy1955}, a unique, satisfactory
answer is still lacking. One is still offered with diverging
options. If one starts from Minkowski's energy tensor
\cite{ref:Minkowski1908}, like Jauch and Watson did, one is
presented with photons endowed with an unpalatable {\it spacelike}
momentum-energy four-vector. If one starts instead from Abraham's
energy tensor \cite{ref:Abraham1909}, as both strong theoretical
arguments \cite{ref:Gordon1923} and sound experimental evidence
\cite{ref:Walker1975a,ref:Walker1975b} might suggest to do, one
reaches the conclusion that photons in dielectric matter can not
be defined, because the energy operator and the momentum operator
have no common eigenvectors; furthermore, both the field energy
and the field momentum are not constant in time
\cite{ref:Nagy1955}. This seemingly odd result is instead quite
reasonable, for Abraham's tensor density has a nonvanishing
four-divergence also in the absence of charges and currents,
thereby entailing a continuous exchange of energy and momentum
with the dielectric medium. Only by extracting from Abraham's
tensor a so called radiation tensor \cite{ref:MN1955} with
vanishing four-divergence did Nagy succeed \cite{ref:Nagy1955} in
producing photons in dielectric matter endowed with the expected
timelike momentum-energy four-vector.\par Since the subsequent
works, quoted e.g. in ref. \cite{ref:Miglietta1998}, have not led
to a solution of the problem at issue, the reader with an
ingenuous-realistic epistemological inclination may find
interesting to learn what ``die Physik der Modelle'' has to say
about photons in matter. For this reader we shall recall the
existence of two very important results obtained many years ago.
One was found by Whittaker et al. \cite{ref:Whittaker1933} and by
Synge \cite{ref:Synge1935}, the second one is already contained in
the seminal paper by Gordon \cite{ref:Gordon1923} and was later
retrieved and extended \cite{ref:Pham Mau Quan1957} by Pham Mau
Quan in a much simpler way through the theory of characteristics;
both are now practically forgotten\footnote{There has been a
recent upsurge of interest in Gordon's effective metric. See e.g.
\cite{ref:Novello2001} and references contained therein.}. The joint
use of these achievements of the past allows one to draw conclusions
about the very existence of photons in isotropic dielectrics.\par
\section{Synge's photons in the vacuum of general relativity}
``Die Physik der Modelle'' has no room for such esoteric ideas as
the wave-particle dualism, since things are supposed to exists in
spacetime in a way totally independent of any act of measurement.
Nothing however forbids to think that one and the same physical
reality may be accounted for by more than one theoretical model.
Therefore it is totally legitimate to check whether a given
classical theory, let us say general relativity, can contain both
a wave description and a particle description of the photon, and
to investigate how the characteristic quantities of both
descriptions are mutually related. This is just what the authors
of refs. \cite{ref:Whittaker1933} and \cite{ref:Synge1935} have
done for the photons in the vacuum of general relativity. Due to
its masterful clearness, we shall report here Synge's account {\it
in extenso}.\par If the atoms are treated as points, it is
necessary that the photon be pictured as something capable of
being emitted and absorbed by points. There is a venerable line of
thought that leads to assume that the world-line of a photon is a
null geodesic. In fact the forerunners of the photons, i.e. the
light rays of geometrical optics, were postulated to be null
geodesics of the spacetime metric $g_{ik}$ already in 1917 by
Hilbert \cite{ref:Hilbert1917}. The postulate was then shown to be
a mere consequence of the electromagnetic equations in a
gravitational field by von Laue \cite{ref:von Laue1920}; the same
result was later reproduced by Whittaker in a two-page note
\cite{ref:Whittaker1927} that relies on the theory of
characteristics. Synge was thus led to assume that the world line
of a photon satisfies the equations
\begin{equation}\label{2.1}
\frac{\delta}{\delta u}\frac{dx^i}{du}=0,
~~~g_{ik}\frac{dx^i}{du}\frac{dx^k}{du}=0,
\end{equation}
where ${\delta}/{\delta u}$ stands for the absolute derivative
with respect to the special parameter $u$.\par If the photon can
be modeled as a point particle, we shall attribute to it a
momentum-energy vector $M^i$ that is tangent to the photon's world
line and is parallelly propagated along it. One therefore poses:
\begin{equation}\label{2.2}
M^i=\theta\frac{dx^i}{du},~~~\frac{\delta M^i}{\delta u}=0,
\end{equation}
where $\theta$, assumed to be positive, happens to be a constant
thanks to (\ref{2.1}). Since a photon is supposed to travel with
the speed of light, it must have zero proper mass. Therefore it
must be
\begin{equation}\label{2.3}
M_iM^i=0;
\end{equation}
the index of $M^i$ has been lowered with the metric $g_{ik}$; we
choose by convention that the Minkowski form to which the latter
can be transformed locally is $\eta_{ik}\equiv diag({1,1,1,-1})$.
We shall model an atom like a timelike point particle endowed with
the momentum-energy vector
\begin{equation}\label{2.4}
m_0 u^i=m_0\frac{dx^i}{ds},~~~ds^2\equiv-g_{ik}dx^idx^k;
\end{equation}
$m_0$ is the atom's rest mass or rest energy, while $u^i$ is its
four-velocity. The energy collected by an atom through the
absorption of a photon can be calculated by postulating that the
absorption process is ruled by the conservation law:
\begin{equation}\label{2.5}
m'_0 {u'}^i=m_0 u^i+M^i,
\end{equation}
where $m'_0$ is the rest mass, and ${u'}^i$  is the four-velocity
of the atom after the absorption of the photon. By rewriting the
conservation law in covariant form and by multiplying the two
forms term by term one finds
\begin{equation}\label{2.6}
-{m'}_0^2=-m_0^2+2m_0M_iu^i,
\end{equation}
i.e.
\begin{equation}\label{2.7}
\frac{{m'}_0^2-m_0^2}{2m_0}=-M_iu^i.
\end{equation}
For optical processes $({m'}_0-m_0)/{m_0}$ is very small with
respect to unity, and this fact suggests defining the energy $E$
of a photon with momentum-energy vector $M^i$ {\it relative} to a
pointlike ``atom'' endowed with four-velocity $u^i$ as
\begin{equation}\label{2.8}
E=-M_iu^i.
\end{equation}
But another, equally time-honoured tradition leads to think of a
photon as a wavelike phenomenon and to draw, with Synge, the
spacetime diagram of Figure 1. In the wave model the photon is not
pictured by just one null world-line, but by the two-dimensional
ribbon $A_0B_0BA$, delimited by the null geodesics $A_0A$, $B_0B$,
and by the line elements $ds_0=A_0B_0$, ${ds=AB}$, taken on the
world-lines of the emitting and of the absorbing atom
respectively.\par
\begin{figure}[h]
\includegraphics{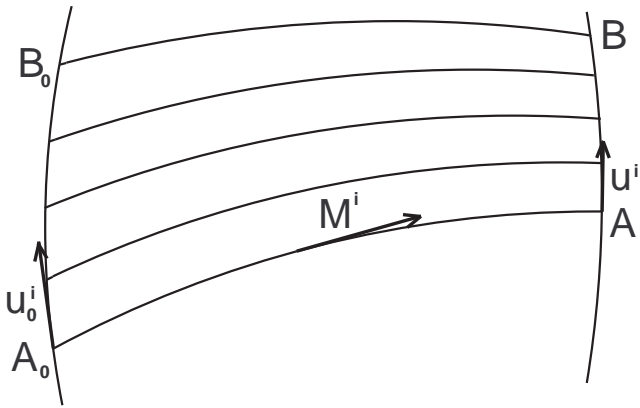}
\caption{Spacetime diagram of Synge's ``photon''.}
\end{figure}
\noindent To each null world-line within the ribbon it corresponds
a phase of the wave process. The null lines drawn in the figure
have, say, zero phase. If these lines are $n+1$ in number, it
means that the emission process entails $n$ periods, and $n$ too
shall be the number of the periods occurring in the absorption
process. Therefore the frequency of emission, defined as the
number of periods per unit proper time along the world-line of the
emitting atom, shall read (with an ugly but otherwise innocuous
licence in the notation)
\begin{equation}\label{2.9}
\nu_0=n/ds_0,
\end{equation}
while the frequency of absorption, defined with respect to the
world line of the absorbing atom, turns out to be
\begin{equation}\label{2.10}
\nu=n/ds.
\end{equation}
By comparing these two equations one finds
\begin{equation}\label{2.11}
\nu_0ds_0=\nu ds,
\end{equation}
i.e. the relation governing the spectral shift in a gravitational
field, whose quite general, geometrical derivation was first given
in 1923 by Lanczos \cite{ref:Lanczos1923}.\par The special
parameter $u$ must be chosen on each of the null geodesics of the
ribbon $A_0B_0BA$. Since each of the $u$'s is given up to a linear
transformation $u'=au+b$, where $a$ and $b$ are constants that can
be arbitrarily fixed on each null geodesic, we can freely assume
that these parameters have a single starting value on $A_0B_0$ and
a single terminal value on $AB$. If $v$ is another parameter,
taken to be constant along the single null geodesic, the 2-space
of the ribbon can be described by the equation
\begin{equation}\label{2.12}
x^i=x^i(u,v),
\end{equation}
while of course the partial differential equation
\begin{equation}\label{2.13}
g_{ik}\frac{\partial x^i}{\partial u}
\frac{\partial x^k}{\partial u}=0
\end{equation}
needs to be satisfied. We are now equipped for considering further
the particle model of the photon and for studying the relation
between the energy release at the atom on the left side of the
figure and the energy intake occurring at the atom on the right
side. If $M^i$ is the momentum-energy vector of the photon, we can
write
\begin{equation}\label{2.14}
\frac{\partial}{\partial u}
\left(M_i\frac{\partial x^i}{\partial v}\right)
=\frac{\delta M_i}{\delta u}\frac{\partial x^i}{\partial v}
+M_i\frac{\delta}{\delta u}\frac{\partial x^i}{\partial v}.
\end{equation}
We have also
\begin{equation}\label{2.15}
\frac{\delta}{\delta u}\frac{\partial x^i}{\partial v}
\equiv \frac{{\partial}^2 x^i}{\partial u\partial v}
+\Gamma^i_{kl}\frac{\partial x^k}{\partial v}
\frac{\partial x^l}{\partial u}
=\frac{\delta}{\delta v}\frac{\partial x^i}{\partial u},
\end{equation}
where $\Gamma^i_{kl}$ is the Christoffel connection built with
$g_{ik}$. Since $\delta M_i/\delta u=0$, the right hand side of
(\ref{2.14}) can be rewritten as
\begin{equation}\label{2.16}
M_i\frac{\delta}{\delta u}\frac{\partial x^i}{\partial v}
=\theta g_{ik}\frac{\partial x^k}{\partial u}
\frac{\delta}{\delta v}\frac{\partial x^i}{\partial u}
=\frac{1}{2}\theta\frac{\partial}{\partial v}
\left(g_{ik}\frac{\partial x^i}{\partial u}
\frac{\partial x^k}{\partial u}\right)=0.
\end{equation}
Hence one finds
\begin{equation}\label{2.17}
\frac{\partial}{\partial u}
\left(M_i\frac{\partial x^i}{\partial v}\right)=0,
\end{equation}
and, with reference to Figure 1:
\begin{equation}\label{2.18}
\left(M_i\frac{\partial x^i}{\partial v}\right)_{A_0}
=\left(M_i\frac{\partial x^i}{\partial v}\right)_A.
\end{equation}
Let $u^i_0$, $u^i$ be the four-velocities of the atoms at $A_0$
and at $A$ respectively. If $dv$ is the infinitesimal increment of
the parameter $v$ when going from the null geodesic $A_0A$ to the
neighbouring one $B_0B$, we can write:
\begin{equation}\label{2.19}
\left(\frac{\partial x^i}{\partial v}\right)_{A_0}dv
=u^i_0ds_0,~~~
\left(\frac{\partial x^i}{\partial v}\right)_Adv
=u^ids,
\end{equation}
hence from (\ref{2.18}) we get
\begin{equation}\label{2.20}
\left(M_i\right)_{A_0}u^i_0ds_0=\left(M_i\right)_Au^ids.
\end{equation}
But the definition (\ref{2.8}) of the energy of a photon absorbed
(or emitted) by an atom allows to rewrite the last equation as
just
\begin{equation}\label{2.21}
E_0ds_0=Eds.
\end{equation}
This is a really momentous result: if we divide term by term this
equation, derived by considering the photon as a particle, and
equation (\ref{2.11}), that was obtained by considering the photon
as a wave, we get eventually that it must be:
\begin{equation}\label{2.22}
\frac{E_0}{\nu_0}=\frac{E}{\nu}.
\end{equation}
Therefore, general relativity  contains both a wave model and a
particle model of the photon {\it in vacuo}, and the relation
between the two models is such that when a photon is emitted by
one atom and absorbed by another one in presence of a
gravitational field, the ratio energy/frequency is the same for
emission and for absorption. This ratio is independent of the
behaviour of the gravitational field and of the state of motion of
the two atoms.

\section{Gordon's reductio ad vacuum of the electromagnetic theory
for a homogeneous, isotropic dielectric}

In 1923 Walter Gordon showed \cite{ref:Gordon1923} that Maxwell's
equations for matter that is homogeneous and isotropic when
considered in its rest frame can be rewritten as Maxwell's
equations {\it in vacuo}, provided that we adopt, in writing those
equations, not the true spacetime metric $g_{ik}$, but the
effective metric
\begin{equation}\label{3.1}
\sigma_{ik}=g_{ik}+\left(1-{\frac{1}{\epsilon\mu}}\right)u_{i}u_{k}
\end{equation}
in which $\epsilon$, $\mu$ are respectively the dielectric
constant and the magnetic permeability of the dielectric medium,
while $u^i$ is its four-velocity. This is a remarkable finding
with far reaching consequences, and we shall spend some words to
show how it comes about. By following the established convention
\cite{ref:Post1962}, we represent the electric displacement and
the magnetic field by the antisymmetric, contravariant tensor
density ${\bf H}^{ik}$, while the electric field and the magnetic
induction are accounted for by the skew, covariant tensor
$F_{ik}$. With these geometrical objects we define the
four-vectors:
\begin{equation}\label{3.2}
F_{i}=F_{ik}u^{k},~~~H_{i}=H_{ik}u^{k},
\end{equation}
where $u^i$ is the four-velocity of matter. In general relativity
a linear electromagnetic medium can be told to be homogeneous and
isotropic in its rest frame if its constitutive equation reads
\begin{equation}\label{3.3}
\mu{H^{ik}}=F^{ik}+(\epsilon\mu-1)(u^{i}F^{k}-u^{k}F^{i}).
\end{equation}
Gordon noticed that (\ref{3.3}) can be rewritten as
\begin{equation}\label{3.5}
\mu{H^{ik}}=\left[g^{ir}-(\epsilon\mu-1)u^{i}u^{r}\right]
\left[g^{ks}-(\epsilon\mu-1)u^{k}u^{s}\right]F_{rs},
\end{equation}
and, since the contravariant form of the effective metric tensor
(\ref{3.1}) is
\begin{equation}\label{3.6}
\sigma^{ik}=g^{ik}-(\epsilon\mu-1)u^{i}u^{k},
\end{equation}
the constitutive equation can be rewritten as
\begin{equation}\label{3.7}
\mu{\bf H}^{ik}=\sqrt{g}\sigma^{ir}\sigma^{ks}F_{rs},
\end{equation}
where $g\equiv-\det(g_{ik})$. With some simple algebra one finds
\cite{ref:Gordon1923} that
\begin{equation}\label{3.9}
\sigma=\frac{g}{\epsilon\mu},
\end{equation}
where $\sigma\equiv-\det(\sigma_{ik})$. Hence (\ref{3.3}) can be
eventually rewritten as
\begin{equation}\label{3.10}
{\bf H}^{ik}=\sqrt{\frac{\epsilon}{\mu}}
\sqrt{\sigma}\sigma^{ir}\sigma^{ks}F_{rs}.
\end{equation}
Therefore, apart from the constant factor $\sqrt{\epsilon/\mu}$,
the constitutive equation for this dielectric medium is just the
same as the one occurring for a general relativistic vacuum. All
the equations and all the theorems that hold for electromagnetism
in the vacuum case will apply also to our medium, provided that
$\sigma_{ik}$ be substituted for $g_{ik}$ in the original vacuum
equations. In particular, we see that the results \cite{ref:von
Laue1920,ref:Whittaker1927} found by von Laue and by Whittaker for
the propagation of light rays in the vacuum of general relativity
immediately apply to our dielectric medium: in the limit of
geometrical optics, the light rays in a dielectric that is
homogeneous and isotropic in its local rest frame shall be
\cite{ref:Gordon1923} the null geodesics of an ``effective
spacetime'' endowed with the metric (\ref{3.1}).\par Gordon's idea
of the effective metric was later resumed by Pham Mau Quan
\cite{ref:Pham Mau Quan1957}. By availing of the theory of
characteristics, he could extend Gordon's just quoted result to
dielectrics that are isotropic, but not homogeneous, since
$\epsilon$ and $\mu$ are assigned functions of the coordinates
$x^i$.
\section{The relation between the wave and the particle model of
the photon in an isotropic dielectric} Thanks to the findings of
Whittaker, Synge, Gordon and Pham Mau Quan recalled in the two
previous sections, we can build both a wave and a particle model
of the photon in an isotropic dielectric and investigate how the
two models are mutually related in the new situation.\par The wave
model of the photon in an isotropic dielectric, i.e. in a medium
with the constitutive equation (\ref{3.3}), where $\epsilon$ and
$\mu$ have an assigned dependence on $x^i$, needs no explanation.
The spacetime diagram of Figure 1 can be drawn as it stands also
in the case of the isotropic dielectric, and we find that the
relation between the frequencies of the photon measured on the
world lines of the emitting and of the absorbing atoms is still
given by equation (\ref{2.11}). Let us only emphasize that the
line elements $ds_0$ and $ds$ measure the proper time of the
pointlike atoms in the actual spacetime and are therefore defined
by the true spacetime metric $g_{ik}$.\par The particle model of
the photon entails instead important changes, that are however
both mandatory and natural, if one keeps in mind the equivalence
between the propagation in the dielectric and the propagation in a
general relativistic vacuum with the effective metric
$\sigma_{ik}$ given by equation (\ref{3.1}). We know from
\cite{ref:Pham Mau Quan1957} that the light rays in the dielectric
are null geodesics with respect to that metric. The definition of
these geodesics is no longer provided by equation (\ref{2.1}). It
reads instead:
\begin{equation}\label{4.1}
\frac{^{\sigma}\delta}{\delta u}\frac{dx^i}{du} \equiv
\frac{d^2x^i}{du^2}+\Sigma^i_{kl}\frac{dx^k}{du}\frac{dx^l}{du}=0
\end{equation}
{\it cum}
\begin{equation}\label{4.2}
\sigma_{ik}\frac{dx^i}{du}\frac{dx^k}{du}=0;
\end{equation}
we shall henceforth denote with ${^{\sigma}\delta}/{\delta u}$ the
absolute derivative performed with the affine connection
\begin{equation}\label{4.3}
\Sigma^i_{kl}=\frac{1}{2}\sigma^{im}\left(\sigma_{mk,l}+\sigma_{ml,k}
-\sigma_{kl,m}\right);
\end{equation}
$u$ is again a special parameter. If the photon in the dielectric
can be modeled as a point particle, we shall attribute to it a
momentum-energy vector $M^i$ tangent to its world line, defined by
(\ref{4.1}) and (\ref{4.2}); the equivalence with the vacuum case
forces us to assume that $M^i$ is parallelly propagated along that
world line. One therefore still poses
\begin{equation}\label{4.4}
M^i=\theta\frac{dx^i}{du},
\end{equation}
but now, instead of $\delta M^i/\delta u=0$, one shall write
\begin{equation}\label{4.5}
\frac{^\sigma\delta M^i}{\delta u}=0.
\end{equation}
Thanks to (\ref{4.1}) the scalar $\theta$, assumed to be positive,
happens again to be a constant. This photon travels with the speed
of light in the spacetime with the metric $\sigma_{ik}$; hence in
this ``effective spacetime'' it must exhibit zero proper mass.
Therefore it must be:
\begin{equation}\label{4.6}
M_{(i)}M^i\equiv\sigma_{ik}M^iM^k=0;
\end{equation}
we adopt henceforth Gordon's convention of enclosing within round
parentheses the indices moved with the effective metric
$\sigma_{ik}$. Let us now look at this photon in the true
spacetime, the one with $g_{ik}$ as metric tensor. Equation
(\ref{4.6}) can be rewritten as
\begin{eqnarray}\label{4.7}
M_{(i)}M^i=\left[g_{ik}+
\left(1-\frac{1}{\epsilon\mu}\right)u_iu_k\right]M^iM^k\\\nonumber
=M_iM^i+\left(1-\frac{1}{\epsilon\mu}\right)\left(u_iM^i\right)^2=0,
\end{eqnarray}
When measured with the metric $g_{ik}$ of the true spacetime the
momentum-energy vector $M^i$ will not be a null one, and with some
trepidation we now check whether it is spacelike, as it occurs
with the photon obtained by Jauch and Watson in their
phenomenological quantum electrodynamics \cite{ref:Jauch1948}, or
whether it is timelike, as Nagy instead found \cite{ref:Nagy1955},
and as it is required by the consistency of the particle model.
Equation (\ref{4.7}) says that
\begin{equation}\label{4.8}
M_iM^i=\left(\frac{1}{\epsilon\mu}-1\right)\left(u_iM^i\right)^2,
\end{equation}
and since $\epsilon\mu>1$, one finds that $M_iM^i<0$. Therefore,
with our sign convention for the metric, the momentum-energy
vector of the photon happens to be timelike; the particle model is
a consistent one, and the photon in the dielectric behaves like an
ordinary massive point particle, for which a rest frame
exists.\par As a consequence we can repeat here, with the due
changes, the argument leading to equation (\ref{2.21}). We start
again from the conservation equation (\ref{2.5}) that rules the
absorption process of a photon by an atom; we shall assume that it
holds, in unaltered form, also in the dielectric. However, if we
lower the indices of (\ref{2.5}) with $g_{ik}$ and multiply term
by term the contravariant and the covariant form of the
conservation law we do not get (\ref{2.6}), since $M_iM^i$ no
longer vanishes. One gets instead
\begin{equation}\label{4.9}
-{m'}_0^2=-m_0^2+2m_0M_iu^i
+\left(\frac{1}{\epsilon\mu}-1\right)\left(u_iM^i\right)^2;
\end{equation}
however for optical processes the last term is negligible with
respect to the other ones. The form (\ref{2.8}) can still be used
to define, with the same degree of approximation as in the vacuum
case, the energy $E$ of a photon with momentum-energy vector $M^i$
{\it relative} to a pointlike ``atom'' endowed with four-velocity
$u^i$. We can avail of Figure 1 also for dealing with the photon
as a particle, and we retain the parametrisation of the ``ribbon''
$A_0B_0BA$ exactly as it stands in the vacuum case. Since now
equation (\ref{4.5}) is substituted for the second equation
(\ref{2.2}), let us consider the quantity
\begin{equation}\label{4.10}
\frac{\partial}{\partial u}
\left(M_{(i)}\frac{\partial x^i}{\partial v}\right)
=\frac{^\sigma\delta M_{(i)}}
{\delta u}\frac{\partial x^i}{\partial v}
+M_{(i)}\frac{^\sigma\delta}
{\delta u}\frac{\partial x^i}{\partial v};
\end{equation}
the first term at the right hand side vanishes due to (\ref{4.5}).
We have also
\begin{equation}\label{4.11}
\frac{^\sigma\delta}{\delta u}\frac{\partial x^i}{\partial v}
\equiv \frac{{\partial}^2 x^i}{\partial u\partial v}
+\Sigma^i_{kl}\frac{\partial x^k}{\partial v}
\frac{\partial x^l}{\partial u}
=\frac{^\sigma\delta}{\delta v}\frac{\partial x^i}{\partial u},
\end{equation}
hence we obtain
\begin{equation}\label{4.12}
M_{(i)}\frac{^\sigma\delta}{\delta u}\frac{\partial x^i}{\partial v}
=\theta \sigma_{ik}\frac{\partial x^k}{\partial u}
\frac{^\sigma\delta}{\delta v}\frac{\partial x^i}{\partial u}
=\frac{1}{2}\theta\frac{\partial}{\partial v}
\left(\sigma_{ik}\frac{\partial x^i}{\partial u}
\frac{\partial x^k}{\partial u}\right)=0
\end{equation}
due to (\ref{4.2}). Therefore
\begin{equation}\label{4.13}
\frac{\partial}{\partial u} \left(M_{(i)}\frac{\partial x^i}
{\partial v}\right)=0
\end{equation}
and, again with reference to Figure 1:
\begin{equation}\label{4.14}
\left(M_{(i)}\frac{\partial x^i}{\partial v}\right)_{A_0}
=\left(M_{(i)}\frac{\partial x^i}{\partial v}\right)_A.
\end{equation}
The same argument as in Section 2 leads to write now
\begin{equation}\label{4.15}
\left(M_{(i)}u^i\right)_{A_0}ds_0=\left(M_{(i)}u^i\right)_Ads.
\end{equation}
Since
\begin{eqnarray}\label{4.16}
M_{(i)}u^i\equiv\sigma_{ik}M^iu^k
=\left[g_{ik}+
\left(1-\frac{1}{\epsilon\mu}\right)u_iu_k\right]M^iu^k\\\nonumber
=M_iu^i-\left(1-\frac{1}{\epsilon\mu}\right)M_iu^i
=\frac{M_iu^i}{\epsilon\mu},
\end{eqnarray}
instead of (\ref{2.20}) one finds
\begin{equation}\label{4.17}
\left(\frac{M_iu^i}{\epsilon\mu}\right)_{A_0}ds_0
=\left(\frac{M_iu^i}{\epsilon\mu}\right)_Ads,
\end{equation}
hence one eventually obtains that the ratio, say, between the
energy emitted by the atom on the left side of Figure 1 and the energy
absorbed by the atom on the right side is no longer given by equation
(\ref{2.21}), since now
\begin{equation}\label{4.18}
\left(\frac{E}{\epsilon\mu}\right)_{A_0}ds_0
=\left(\frac{E}{\epsilon\mu}\right)_Ads,
\end{equation}
i.e. the ratio depends on the values that $\epsilon$ and  $\mu$
happen to assume at the spacetime location of the emitting and of
the absorbing atom respectively. Since, however, the relation
(\ref{2.11}) between the frequencies does not undergo any change
when going from the vacuum case to the case of the isotropic
dielectric, instead of the vacuum equation (\ref{2.22}) we get
\begin{equation}\label{4.19}
\left(\frac{E}{\nu\epsilon\mu}\right)_{A_0}
=\left(\frac{E}{\nu\epsilon\mu}\right)_A.
\end{equation}
\vbox to 1cm{}
\section{Conclusion} The last equation summarizes one thing
that general relativity has to say about photons in isotropic
dielectric matter, provided that its ``Physik der Modelle'' is not
declared {\it a priori} incompetent to deal with such a definitely
quantum affair. In the case of the general relativistic vacuum,
investigated by Whittaker et al. \cite{ref:Whittaker1933} and by
Synge \cite{ref:Synge1935}, the particle and the wave model of the
photon are mutually related in a way that, at least for optical
photons, has an important point of agreement with the quantum
theory of radiation. One should not be as naive as to identify
concepts stemming from so different theoretical constructions like
general relativity and quantum field theory just because these
concepts have been christened with the same name in both theories.
But also with this proviso it is indeed remarkable that in general
relativity the ratio of the two invariant quantities $E$ and
$\nu$, associated to the photon as shown in Section 2, has at the
absorption just the same value that it exhibits at the emission.
This ratio is totally independent of the way the photon travels
from the emitting to the absorbing atom, and is also independent
of the state of motion of these atoms. A theoretician that still
has a penchant for an ingenuous-realistic epistemology can feel
enticed by these findings into taking very seriously the concept
of photon {\it in vacuo}, in keeping with Einstein's famous
complaint\footnote{ Letter to Michele Besso of December 12, 1951:
``The whole fifty years of conscious brooding have not brought me
nearer to the answer to the question `What are light quanta?'.
Nowadays every scalawag believes that he knows what they are, but
he deceives himself''\cite{ref:Speziali1972}. English translation
by J. Stachel \cite{ref:Stachel1986}.}.\par The same theoretician
will draw from the result of Section 4, expressed in equation
(\ref{4.19}), the following hints. If the dielectric is
homogeneous and isotropic, it is likely that the program
\cite{ref:Miglietta1998} of deriving an effective Hamiltonian from
a fully quantum description of the whole system consisting of
matter plus field will be worth the inordinate effort it
presumably requires. In fact the classical wave and the classical
particle models do exist and show that in such a medium the ratio
of the invariants $E$ and $\nu$ will be the same at the emission
and at the absorption, just as it happens {\it in vacuo}. For a
homogeneous, isotropic dielectric the very existence of
``phenomenological photons'' derived by quantization from the
effective Hamiltonian mentioned above is therefore an expected
occurrence, since a classical simile is already known to
exist.\par If the dielectric is not homogeneous, however, the same
equation (\ref{4.19}) shows that the constancy of the ratio
between $E$ and $\nu$ is no longer ensured. Therefore, for such a
medium, a classical simile of the quantum theoretical photon
cannot be found. One should remind of this circumstance before
waving the magic wand of quantisation over an effective
Hamiltonian for an inhomogeneous dielectric, be it derived from
the quantum electrodynamics of vacuum, or just drawn with some
argument from the classical electrodynamics of continua.

\end{document}